\documentclass[mauscript]{aastex}
\usepackage{amssymb}
\usepackage{pifont}
\usepackage{graphicx}
\RequirePackage{longtable}

\shortauthors{Yan et al.}

\begin{document}

\title{Phase Evolution of the Crab Pulsar between Radio and X-ray}

\author{L.L. Yan\altaffilmark{1,2,3}, M.Y. Ge\altaffilmark{1},
J.P. Yuan\altaffilmark{4}, S.J. Zheng\altaffilmark{1},
F.J. Lu\altaffilmark{1}, Y. L. Tuo\altaffilmark{1,2}, H. Tong\altaffilmark{4},
S.N. Zhang\altaffilmark{1}, Y. Lu\altaffilmark{1}, J.L. Han\altaffilmark{5}, Y.J. Du\altaffilmark{3}}

\affil{$^1$Key Laboratory of Particle Astrophysics, Institute of High Energy Physics,
Chinese Academy of Sciences, Beijing 100049, China; yanlinli@ihep.ac.cn}
\affil{$^2$University of Chinese Academy of Sciences, Beijing 100049, China.}
\affil{$^3$Qian Xuesen Laboratory of Space Technology, No. 104, Youyi Road,
Haidian District, Beijing 100094, China}
\affil{$^4$Xinjiang Astronomical Observatory, Chinese Academy of Sciences,
Urumqi, Xinjiang 830011, China}
\affil{$^5$National Astronomical Observatory, Chinese Academy of Sciences,
Jia 20 Datun Road, Beijing 100012, China}

%\published{Published in {\apj} on August 18, 2017}
%\submitjournal{AASJournal name}

\begin{abstract}

We study the X-ray phases of the Crab pulsar utilizing the 11-year
observations from the {\sl Rossi X-ray Timing Explorer}, 6-year radio
observations from the Nanshan Telescope, and the ephemeris from Jodrell
Bank Observatory. It is found that the X-ray phases in different energy
bands and the radio phases from Nanshan Telescope show similar
behaviors, including long-time evolution and short-time variations.
Such strong correlations between the X-ray and radio phases imply
that the radio and X-ray timing noises are both generated from the
pulsar spin that cannot be well described by the
the monthly ephemeris from the Jodrell Bank observatory. When using the
Nanshan phases as references to study the X-ray timing noise, it has
a significantly smaller variation amplitude and shows no long-time
evolution, with a change rate of $(-1.1\pm1.1)\times10^{-7}$\,
periods per day. These results show that the distance of the X-ray
and radio emission regions on the Crab pulsar has no detectable secular
change, and it is unlikely that timing-noises resulted from any unique
physical processes in the radio or X-ray emitting regions. The similar
behaviors of the X-ray and radio timing noises also imply that the
variation of the interstellar medium is not the origin of the Crab
pulsar's timing noises, which is consistent with the results obtained from
the multi-frequency radio observations of PSR B1540$-$06.

\end{abstract}


\section{Introduction}
Pulsars are famous for their rotation stability and highly
repeatable pulse shapes. However, when examining pulsar's
periodicity with high precision, there appear to be two main types
of irregularities, namely glitch and timing noise. The origin
of timing noise remains controversial in spite of years of
studies. Models proposed to explain the timing noises of pulsars
include the random process \citep{Cordes(1980)}, unmodeled planetary
companions \citep{Cordes(1993)}, the free precession \citep{Stairs et al.(2000)},
and the interstellar medium (ISM; e.g. \cite{You et al.(2007)}).
\cite{Hobbs(2010)} analyzed the timing properties of 366 pulsars
in detail. Their study shows that timing noise is widespread in
pulsars and cannot be explained using a simple random walk in the
observed rotational parameters. The timing residuals of
PSR B1540$-$06 are consistent at different radio frequencies, which
implies that its timing noise is not caused by ISM \citep{Hobbs(2010)}.
The underlying physical processes that cause timing noise
are still unclear.

Among all the pulsars, the Crab pulsar is probably the most suitable
source to study the origin of timing noise for its frequent spin
irregularities and abundant observational data. This pulsar has been
comprehensively studied in almost all wavelength bands from radio
to very high energy $\gamma$-rays. Its pulse profile shows a double-peak
structure in all of these wavebands. Generally, the two dominant
pulses in the radio band are denoted as main pulse (MP) and interpulse
(IP; \cite{Lyne et al.(2013)}), and the two X-ray peaks are denoted
as P1 and P2 \citep{Kuiper et al.(2001)}. Detailed studies show that
the exact pulse morphology varies as a function of photon
energy \citep{Abdo et al.(2010),Ge et al.(2012)} and the positions of
the main peak in different energy bands are not exactly aligned, i.e., the
optical, X-ray and $\gamma$-ray pulses lead to the radio pulses
\citep{Kuiper et al.(2003),Rots(2004),Oosterbroek et al.(2008),Abdo et al.(2010),Molkov(2010)}.
Recently, secular changes of both the radio and the X-ray profiles
were found, though their change rates are different from each other \citep{Lyne et al.(2013),Ge et al.(2016)}.
These changes were attributed to a progressive change in the magnetic
inclination \citep{Lyne et al.(2013),Ge et al.(2016)}, and such magnetic
field variations are also confirmed by another study on red timing
noise \citep{Yi(2015)}.
With seven years of observations, it was found that the X-ray pulse in
2--16 keV leads the radio one by $0.0102\,\pm\,0.0012$ period and increases
with a rate of $(3.3\pm2.0)\times10^{-7}$ period per day \citep{Rots(2004)},
which could also be explained as systematic errors. Given the secular changes
of the radio and X-ray profiles, to study the phase lags between different
energy bands and their variations is important to uncover the origin of
timing noise and the properties of magnetosphere structure.

In this paper, by using the 11-year observations from the {\sl Rossi
X-ray Timing Explorer} ({\sl RXTE}), 6-year radio observations
from Nanshan radio telescope at the Xinjiang Astronomical Observatory \citep{Wang et al.(2001)},
and the monthly renewed ephemeris from the Jodrell Bank Observatory
\citep{Lyne(1993)}, we investigated in detail the timing behaviors
of this pulsar in the X-ray and radio wavebands. First, the phase
comparisons between the Proportional Counter Array (PCA) and the
High Energy X-ray Timing Experiment (HEXTE) on board {\sl RXTE} are
used to estimate the instrumental influences on the phase
determination. Then, the accuracy of the Jodrell Bank ephemeris is
checked by the correlation between the X-ray phases and the radio
phases obtained by the Nanshan radio telescope. Furthermore, the
X-ray phases are corrected by the new phase indicator from Nanshan
radio telescope to study the relationship between the X-ray and radio
timing noises, including the effects of the dispersion measure (DM).

\section{Observations and Data Reduction}

\subsection{Timing Ephemeris from Jodrell Bank}

In this study, the time reference for the radio phases from Nanshan
and X-ray phases from {\sl RXTE} of the Crab pulsar is taken as the
times-of-arrival (TOAs) from Jodrell Bank radio ephemeris (JBE;
\cite{Lyne(1993)}). A 13 m radio telescope at Jodrell Bank monitors
the Crab pulsar daily, offering a radio ephemeris
\footnote{http://www.jb.man.ac.uk/pulsar/crab.html} that is used for
the analyses of {\sl RXTE} and Nanshan data. The ephemeris we
used is in CGRO format, the format required by the {\sl RXTE}
data processing, and it contains the following information:
R.A. and decl. in J2000 coordinates, the first and
last dates for valid parameters, the infinite-frequency geocentric
UTC TOA of a pulse, rotation frequency and its first two derivatives,
the barycentric (TDB) epoch of the spin parameters, and the
root-mean-square radio timing residual.
Because of the uncertainties of the radio receiver system and
the calibration, we add a systematic error of 40\,$\mu$s for phases
calculated from this ephemeris as suggested by \cite{Rots(2004)}. All
errors of the phases in this paper are 1$\sigma$, for both statistical
and systematic errors.

\subsection{{\sl RXTE} Observations and Data Reduction}

The X-ray observations used in this paper were obtained by
both PCA and HEXTE on board the {\sl RXTE}. The PCA instrument is
composed of five Proportional Counter Units with a total photon
collection area of $6500$ cm$^{2}$. Its effective energy range
is 2--60 keV, and the time resolution is about 1\,$\mu$s (in Good Xenon mode;
\cite{Jahoda et al.(2006)}). These properties make PCA an ideal
instrument to study the detailed temporal properties of pulsars.
In this paper, we use the publicly available data in event mode
E\_250us\_128M\_0\_1s. The time resolution of this mode is
about 250\,$\mu$s. HEXTE consists of two independent detector
clusters A and B, and each of them contains four NaI(Tl)/CsI(Na)
scintillation detectors. This instrument is sensitive in 15--250 keV,
with a detection area of $1600$ cm$^{2}$ and a time resolution of
$7.6$\,$\mu$s \citep{Rothschild et al.(1998)}. In its default operation
mode, the field of view of HEXTE, each cluster is switched on and off source
to provide instantaneous background measurements.
The HEXTE data used in this paper are in mode E\_8us\_256\_DX0F.

The PCA and HEXTE data were analyzed by using the FTOOL from
the astronomy software HEASOFT (v6.15). The method of
data reduction and pulse profile calculation is the same as
in \cite{Ge et al.(2016)}, but with the observations selected
a little differently. Only 243 observations between MJD 51955
 (2001 February 15) and 55927 (2012 January 01) are used in this
paper, since from each of these observations high statistical
PCA and HEXTE pulse profiles can be obtained simultaneously. The
 pulse profiles were binned into 1000 phase bins, in energy bands
  2--60 keV for PCA and 15--250 keV for HEXTE.

\subsection{Nanshan Radio Telescope Observations and Data Reduction}

The Nanshan 25 m radio telescope, operated by Xinjiang
Astronomical Observatory, started to observe the Crab pulsar
frequently in 2000 January. As described in \cite{Wang et al.(2001)},
the two hands of circular polarization at 1540 MHz
are fed through a $2\,\times\,128\,\times\,2.5$ MHz analogue
filter bank. The signal is sampled at 1 ms intervals and 1 bit
digitized. Time is stamped by a hydrogen maser calibrated with
the Global Position system. The integration time of each
observation of the Crab pulsar is 16 minutes. After 2010 January,
the data were obtained with digital filter bank 3 (DFB3), which
 was configured to a bandwidth of 0.5 MHz for each sub-channel and
8 bit sampling. The data are folded online with a subintegration
time of 1 minute for AFB and $30$ s for DFB3, and then written to the
disk with 256 bins across the pulse profile for AFB and 512 bins for DFB.

As for the radio observations, the off-line data reduction
is performed in the following three steps by using the
PSRCHIVE package: (1) the data for the Crab pulsar are
de-dispersed using the DM values from the JBE timing results and
then summed to produce a total intensity profile;
(2) local TOAs were determined by correlating the data with
standard pulse profiles of high signal-to-noise ratio, and the
pulse TOAs normally correspond to the peak of the main
pulse; (3) convert TOAs to the solar system barycentre. The
detail data reduction process is the same as
in \cite{Yuan et al.(2010)}. Because the data quality is better
after 2005 May, we only analyzed those from MJD 53500 (2005 May 10) to
55688 (2011 May 07), and finally got 524 TOAs at 1540 MHz.

\section{Phase Calculation and the Linear Fitting Method}

\subsection{Phase Calculation for {\sl RXTE}}
As described in \cite{Ge et al.(2012)}, the two asymmetrical
pulses of the Crab pulsar in the X-ray band could be modeled
by the formula (1) proposed by \cite{Nelson et al.(1970)}.
In this paper, we only fitted the shape of P1 of the X-ray profile to obtain its
peak phase.
\begin{equation}
L(\phi-\phi_{0})=N\frac{1+a(\phi-\phi_{0})+b(\phi-\phi_{0})^{2}}{1+c(\phi-\phi_{0})
+d(\phi-\phi_{0})^{2}}e^{-h*(\phi-\phi_{0})^{2}}+l,
\label{eq:1}
\end{equation}
where $L$ is the intensity at phase $\phi$, $l$ is the baseline of
the light curve, $\phi_0$ is the phase shift, $N$ is the pulse height
of the profile, and $a$, $b$, $c$, $d$ and $h$ the shape coefficients.
The pulse phase is measured in phase units, of range (0,1). We fitted
the shape of P1 with a relatively broad phase window
(--0.055, 0.0355) centered at phase --0.01.

The calculation procedures of phases of P1 and the estimation of
their statistical errors are the same as in \cite{Ge et al.(2012)}.
The X-ray phases with high precision are obtained after the fitting
procedure, and we denote the phases of the X-ray pulse P1 as
$\Phi_{P}$, $\Phi_{H}$ for PCA and HEXTE, respectively.

\subsection{Phase Calculation for Nanshan}
The calculation procedures of the radio phase of MP for Nanshan
($\Phi_{N}$) are as follows: (1) Remove the dispersion effect for TOAs
using the DM values from JBE \citep{Lyne(1993)}. In this
step, we need to reckon the DM values in Nanshan observations
using linear interpolation. The time delay ($t_{\rm{DM}}$)
caused by DM is
\begin{equation}
t_{\rm{DM}}\,=\,D\times\frac{\rm{DM}}{\nu^{2}},
\label{eq:3}
\end{equation}
where $D$ is the dispersion constant, $D=4.1488\times10^3$
$\rm{MHz^2\,pc^{-1}\,cm^{3}\,s}$, and $\nu$ is the centered
frequency, i.e. 1540 MHz \citep{Lyne2012}. Because the
JBE DM values were
not obtained in the same time as the Nanshan observations, the DM
at the time of Nanshan observations were obtained with linear
interpolation. (2) Convert the TOAs from Jodrell Bank and Nanshan
to the TDB time system. (3) Calculate $\phi_{J}$ for Jodrell Bank and
$\Phi_{N}$ relative to $\phi_{J}$ with formula (\ref{eq:2}) and
(\ref{eq:2_1}), respectively.
\begin{equation}
\phi_{J}=f_{0}(T_{J}-T_{0})+ \frac{1}{2}f_{1}(T_{J}-T_{0})^{2}+\frac{1}{6}f_{2}(T_{J}-T_{0})^{3},
\label{eq:2}
\end{equation}
\begin{equation}
\Phi_{N}= mod\;[f_{0}(T_{N}-T_{0})+ \frac{1}{2}f_{1}(T_{N}-T_{0})^{2}+\frac{1}{6}f_{2}(T_{N}-T_{0})^{3} - \phi_{J}\;,1],
\label{eq:2_1}
\end{equation}
where $T_{J}$ and $T_{N}$ are the TOAs in the TDB time system,
from Jodrell Bank and Nanshan, respectively, $f_{0}$, $f_{1}$
and $f_{2}$ are the spin parameters at the reference epoch
$T_{0}$, and mod is to obtain the residual after removing the
integral periods. The final errors of $\Phi_{N}$ are from
the errors of Nanshan TOAs and the systematic uncertainty of
40 $\mu$s as mentioned previously. In order to investigate the
effect of DM on the timing noises, we also calculate the Nanshan
phases without de-dispersion by skipping the first step above,
and denote it as $\Phi_{N0}$. Because of the process ``mod",
i.e. removing the integral periods, the difference of $\Phi_{N0}$
and $\Phi_{N}$ is smaller than one period (see Fig. \ref{fig6}).

\subsection{Methods for Phase Analysis}

\subsubsection{Linear Fitting}
In order to study the phase variations versus time
and the correlations between phases from different data sets,
we fit the data points with a linear function. For the variation
of a parameter versus time, if the slope deviates significantly
from zero, long-term evolutions should exist. For the correlations
between two parameters, the slope can also tell us information about how
these two parameters are correlated, as we will discuss in section 4.

In this paper, the fitting method is the robust linear modeling (RLM)
from the R statistical software package \citep{Feigelson(2012)}, which 
has been used to study both the phase variations versus time and the 
correlations between different phases. The \emph{MASS} (Modern Applied 
Statistics with S) library based on R-language has the \emph{rlm} function 
for RLM. In this function, the fitting is achieved using an iteratively 
reweighted least-square algorithm. Similarly, the linear fitting for 
the phase correlations between different data groups is also achieved 
by this method, as listed in Table 1 and Table 2.

\subsubsection{Correlation Analysis}

The Pearson's correlation coefficient $r$ \citep{Lee Rodgers(1988)} is
a suitable parameter to describe the influence of the timing noise on
$\Phi_{P}$, $\Phi_{H}$  and $\Phi_{N}$ quantitatively. Because the Nanshan
and RXTE observations were not done simultaneously, and the time series 
of $\Phi_{N}$ is serially dependent when checking the autocorrelation 
function, the Nanshan phases at the time of X-ray observations 
($\Phi_{N}^{'}$) are computed by linear interpolation between the 
neighboring $\Phi_N$ values.

\section{Results}

The X-ray phases $\Phi_{P}$ and $\Phi_{H}$ are the phases
of the X-ray main peaks relative to JBE, from
the PCA and HEXTE data respectively. In order to study the
relation of the radio and X-ray phases on both long and short
time scales, we need to check the accuracy and reliability
of these phases first.

\subsection{The X-Ray Phases from PCA and HEXTE and Their Correlation}

\begin{figure}
\begin{center}
\includegraphics[width=0.8\textwidth]{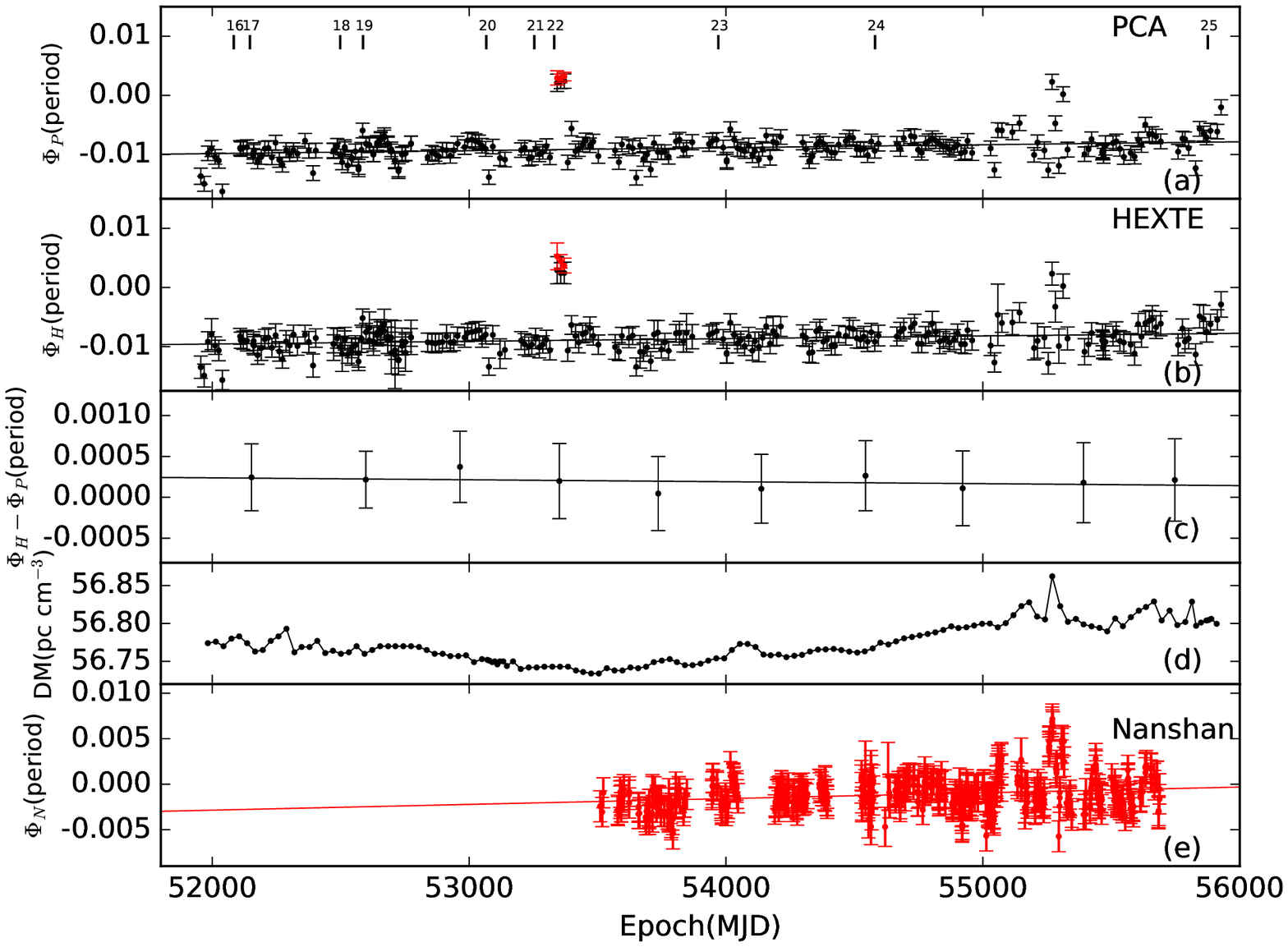}
\caption{X-ray and Nanshan phases of the Crab pulsar from {\sl RXTE},
and DM values in MJD 51955--55927. Panel (a) shows the X-ray phases 
of PCA, 2--60 keV. Panel (b) shows the X-ray phases of HEXTE, 15--250 
keV. The error bars in Panels (a) and (b) include fitting and systematic 
uncertainties. Numbered tick marks in panel (a) indicate the time of 
glitches \citep{Espinoza et al.(2011)}. Black points: the results 
obtained by FASEBIN. Red points: the results obtained by TEMPO2, see 
the details in Section 5.2; Panel (c) shows the phase lags between 
HEXTE and PCA. The data points are averaged over a time of about 400 
days to show them tersely. Panel (d) shows the DM values, they are 
available from the web page http://www.jb.man.ac.uk/pulsar/crab.html 
\citep{Lyne(1993)}. Panel (e) is the Nanshan phases in MJD 53500--55693. 
In panels (a)-(e), the oblique solid lines are the fitting results.
\label{fig1}}
\end{center}
\end{figure}

As shown in Fig. \ref{fig1}, the X-ray phases $\Phi_{P}$ and $\Phi_{H}$
exhibit simultaneous variations on all time scales. Previous work showed
that the X-ray phases from PCA gradually increase with a change rate of
$(3.3\pm2.0)\times10^{-7}$\,period per day (MJD 50129--52941, \cite{Rots(2004)})
or $(6.6\pm1.3)\times10^{-7}$\,period per day (MJD 51955--55142, \cite{Ge et al.(2012)}).
Here we analyze more observations, in a longer time range MJD 51955--55927,
and both PCA and HEXTE showed the same trend with the change rates of
$(5.0\pm0.9)\times10^{-7}$ and $(4.5\pm0.9)\times10^{-7}$\,period
per day, respectively. Besides the increasing trends, $\Phi_{P}$ and $\Phi_{H}$
have two kinds of variations on short time scales, which are the slow
variations (e.g. in MJD 52600--53000 and 55000--55200) and phase jumps
(three points around MJD 53350, corresponding to the JBE in one month).

\begin{figure}
\begin{center}
\includegraphics[angle=0, width=0.8\textwidth]{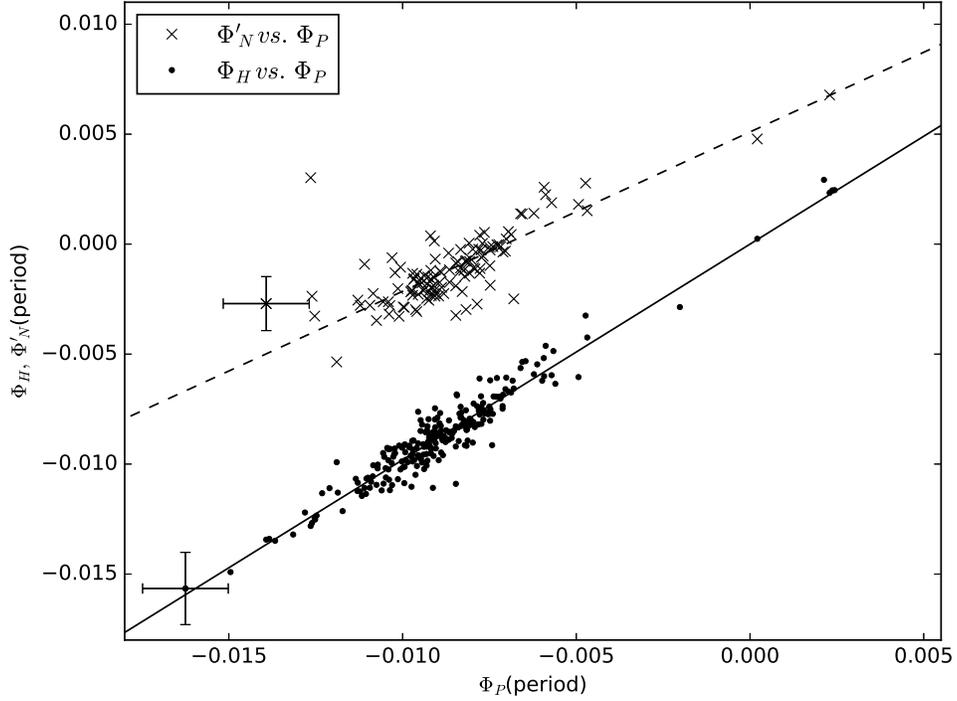}
\caption{Correlations between the X-ray and interpolated Nanshan 
phases. The dots represent the correlation of X-ray phases from PCA 
and HEXTE. The cross marks represent the correlation of X-ray phases 
from PCA and interpolated Nanshan phases in MJD 53500--55693. The 
solid line is the fitting result of dots, while the dash line is 
the fitting result of cross marks. The typical errors are plotted 
to clearly show the correlations.
\label{fig2}}
\end{center}
\end{figure}

Correlation coefficient is calculated to estimate the degree of
correlation between $\Phi_{P}$ and $\Phi_{H}$. As shown in Fig. \ref{fig2} 
and listed in Tab. \ref{table:2}, $\Phi_{H}$ is almost proportional to 
$\Phi_{P}$, with a slope of $0.98\pm0.02$ and the Pearson's coefficient 
$r=0.96$, which means that they vary synchronously with the same amplitude. 
The synchronous variations between $\Phi_{P}$ and $\Phi_{H}$ imply that 
they have the same origin.

\subsection{The Correlation between X-Ray and Nanshan Phases}

\begin{figure}
\begin{center}
\includegraphics[angle=0, width=0.8\textwidth]{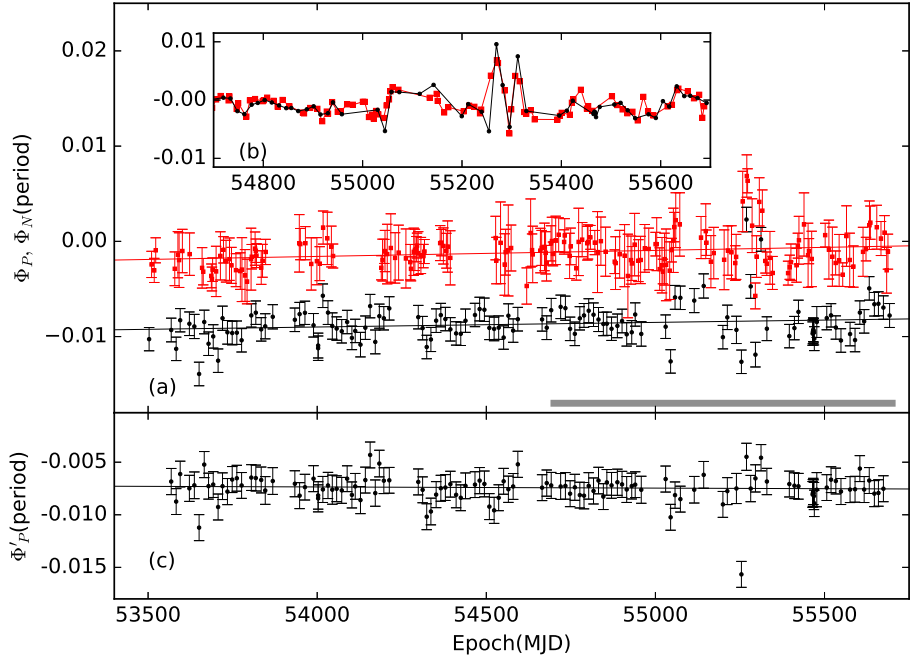}
\caption{X-ray, Nanshan phases and corrected X-ray phases for PCA in 
MJD 53500--55693. Panel (a) shows the variations of X-ray phases from 
PCA (black dot points) and radio phases from Nanshan (red square points). 
The oblique solid lines are the fitting results. The inset (b) shows 
the X-ray and Nanshan phases in MJD 54700--55700 as marked with the gray 
belt in panel (a). The X-ray phases are shifted upward with 0.007 to 
compare them more clearly. The Nanshan phases are averaged in a time 
window of 1.5 days to show them tersely. Panel (c) shows the corrected 
X-ray phases from PCA.
\label{fig3}}
\end{center}
\end{figure}

\begin{figure}
\begin{center}
\includegraphics[angle=0, width=0.8\textwidth]{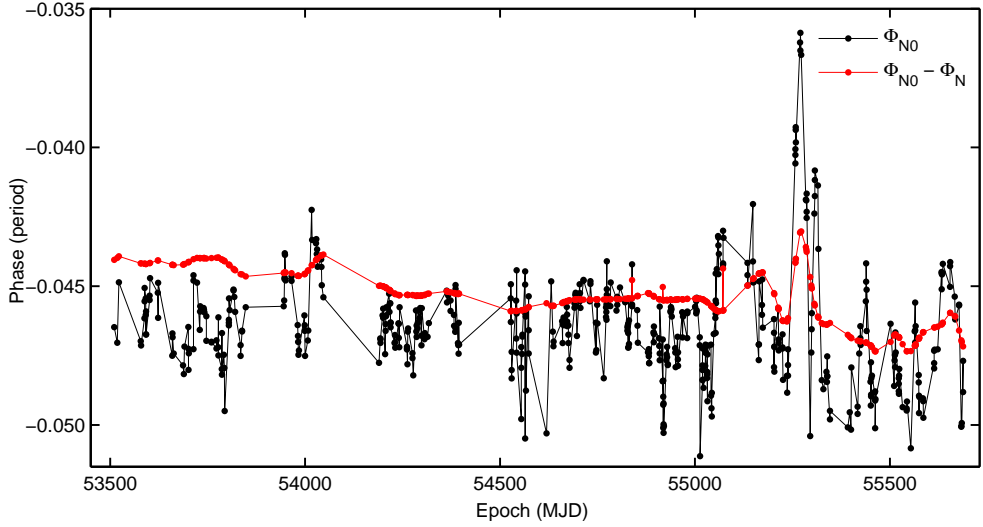}  %%
\caption{The Nanshan phases without de-dispersion and the phase 
differences between them and the Nanshan phase. Black points: the 
Nanshan phases without de-dispersion. Red points: the phase 
difference between Nanshan phases with and without de-dispersion. 
$\Phi_N$ is computed by using the steps $(1)$-$(3)$ in Section 3.2, 
and in order to investigate the effect of DM, $\Phi_{N0}$ is 
computed by skipping the step (1).
\label{fig6}}
\end{center}
\end{figure}

The Nanshan radio phases $\Phi_{N}$ are also obtained by using the
same JBE, and they show variations in different time scales
too, as illustrated in Fig. \ref{fig1}e and Fig. \ref{fig3}a.
Compared with $\Phi_{P}$ in the same time range, $\Phi_{N}$ shows
similar fluctuations, especially in MJD 55000--55200 as in the zoomed in
Fig. \ref{fig3}b. For the secular change, $\Phi_{N}$ increases linearly
with a change rate of $(6.3\pm1.0)\times10^{-7}$\,period per day
in MJD 53500--55688, which is consistent with the change rate $\Phi_{P}$,
$(4.8\pm2.1)\times10^{-7}$\,period per day in the
same time range. These two change rates were obtained in a relatively
short period are also consistent with the results obtained from the
whole time range for $\Phi_{P}$ and $\Phi_{H}$. As shown by the cross
marks in Fig. \ref{fig2}, $\Phi_{N}^{'}$ and $\Phi_{P}$ exhibit a strong
linear correlation, with the Pearson's coefficient $r = 0.78$
(Tab. \ref{table:2}) and a slope of $0.72\pm0.04$. The strong
correlation between $\Phi_{N}^{'}$ and $\Phi_{P}$ means that $\Phi_{N}$
and $\Phi_{P}$ also have a strong correlation.

The fitted slope between $\Phi_{N}$ and $\Phi_{P}$ ($0.72\pm0.04$)
is different from 1, the expected values of $\Phi_{N}$ and $\Phi_{P}$ have
the same variation amplitude and an exactly linear correlation. One may
think that there is some physics behind this. Nonetheless, we realized
that this result probably originated from the data handling process.
Since the X-ray and Nanshan radio observations are carried out in
different times, we obtained the Nanshan phases at the time of X-ray
observations with linear interpolation ($\Phi_{N}^{'}$), so as to study
their correlation. This linear interpolation will reduce the amplitudes
of the radio timing noises, and the larger the amplitude is, the bigger
the fraction of the variation that will be reduced. As a result of this linear
interpolation process, the slope can be smaller than 1.

If the JBE can describe the spin of the Crab pulsar accurately,
$\Phi_{N}$ should be constant over time, because $\Phi_{N}$ is also
inferred from the radio data. However, as given above,
$\Phi_{N}$ has a secular change with a significance of nearly 6.3 $\sigma$,
and both its long-time and short-time variations are similar to the
X-ray ones that are also derived from the JBE. It is very likely that
the temporal behaviors of the Crab pulsar cannot be accurately described
by those spin parameters in the JBE, which also causes the apparent
variation of the X-ray to radio phase lags.

\subsection{The X-Ray Phases by Using the Nanshan Radio Ephemeris}

The simultaneous secular changes and fluctuations of the X-ray and 
Nanshan radio phases imply that they may be caused by the timing noise 
of the Crab pulsar or the inaccuracies in the JBE. In order to further 
check the relations between the X-ray and radio phases, here we use 
the Nanshan phases $\Phi_{N}^{'}$ as the phase references, and obtain 
the corrected X-ray phases $\Phi_{P}^{'}$ from PCA data, which has a 
lower variation amplitude (i.e. standard deviation) 0.0013 compared 
to 0.0020 of $\Phi_{P}$ as shown in Fig. \ref{fig3}c. Moreover, 
$\Phi_{P}^{'}$ keeps almost constant over the time range MJD 53500--55693 
with a change rate of $(-1.1\pm1.1)\times10^{-7}$\,period per day. 
The disappeared secular change of the new X-ray phases suggests that 
the JBE is inaccurate.

\section{Origin of the Phase Variation and Timing Noise}

There are several factors that can result in the variabilities of 
the observed X-ray phases: the instability of the time system, the 
change of the instrument response, the timing noise of the pulsar, 
the inaccuracies of the ephemeris, and the intrinsic variation of 
the X-ray emitting region relative to the radio ones. The effects 
of these factors are discussed in the following.

\subsection{Time System and Instrument Response}

The Mission Operations Center of {\sl RXTE} performs clock calibrations 
several times a day, using the User Spacecraft Clock Calibration System 
method, and the timing accuracy was improved from 4.4 to 2.5\,$\mu$s on 
1997 April 29 \citep{Jahoda et al.(2006)}. Besides, the instrumental 
delay correction for the PCA is 16--20\,$\mu$s and for the HEXTE it is 
0--1 $\mu$s. The barycenter corrections by FTOOL has an accuracy of 
better than 1 $\mu$s and has also subtracted 16 $\mu$s to account for 
the instrumental delay in the 
PCA.\footnote{http://heasarc.gsfc.nasa.gov/docs/xte/abc/time.html}
Therefore, the maximum timing uncertainties is $\sqrt{2.5^2+1^2}+4 =
6.7\,\mu$s, which has a much smaller impact on the timing measurement 
for X-ray photons than the 40\,$\mu$s systematic error from JBE \citep{Rots(2004)}.

If the time systems of {\sl RXTE} and the Nanshan telescope were inaccurate,
wrong timing recorders would be assigned and thus would cause the abnormal
phases. For PCA and HEXTE, the consistent variations might have been caused by the
irregularity of the time system because they use the same time information
from the satellite\footnote{http://heasarc.nasa.gov/docs/xte/time\_news.html}.
However, considering that $\Phi_{P}$ and $\Phi_{N}$ have very similar
variations and they are based on two independent time systems, the
inaccuracy of the time systems cannot account for the observed phase
fluctuations.

The aging of detectors would also have impacts on the timing recorders. 
As the X-ray phase lag of the Crab pulsar evolves with energy
\citep{Molkov(2010), Ge et al.(2012)}, the phase lag will change if
the detection efficiency curve varies due to the instrument aging or 
other factors \citep{Garcia et al.(2014)}. However, the change of the 
response functions of the X-ray instrument cannot explain the correlation 
between $\Phi_{P}$ and $\Phi_{N}$, because the detectors are totally different.

\subsection{Inaccuracies in JBE}

The inaccuracy in the ephemeris have direct impacts on the phase 
calculations for the X-ray and Nanshan phases. The effect of ISM, 
pulsar proper motion, glitches, as well as timing noises could all 
generate inaccurate parameters.

\subsubsection{Effect of the ISM}

Because of the existence of ISM, the arrival time of radio pulses is 
dependent on frequency. Both the mismeasurement of DM and the scattering 
of ISM have an impact on radio observations and phase results.

During MJD 55050--55350, the variation amplitudes of DM are larger than 
in the other time periods, which is apparently consistent with the 
questionable points in this time range. The phase change caused by DM 
could be obtained using $\Phi_{N0}$-$\Phi_{N}$, and its impact on the 
Nanshan phase could be evaluated by comparing it with $\Phi_{N}$. 
However, as shown in Fig. \ref{fig6}, the phase change caused by DM is 
smaller than the phase fluctuation in both X-ray and Nanshan phases in 
MJD 55050--55350, as the DM effect has been removed in the JBE 
\citep{Lyne(1993)}. So, the large fluctuation in the X-ray and Nanshan 
phases in MJD 55050--55350 could not be explained by the DM effects.

As for the scattering of ISM, its influence on TOA could be evaluate by 
the following formula \citep{Lyne2012}.
\begin{equation}
t_{\rm{scatt}}=(\frac{\rm{DM}}{1000})^{3.5}(\frac{400}{\nu_{\rm{MHz}}})^4
\label{eq:5}
\end{equation}
For the Crab pulsar, $\rm{DM}=56.78$ pc cm$^{-3}$, and when 
$\nu_{\rm{MHz}}=1540$ MHz (Nanshan radio band), $t_{\rm{scatt}}\simeq0.2$ $\mu$s. 
In the X-ray band $t_{\rm{scatt}}$ would be much smaller. It is thus clear 
that the influence of ISM scatting on the Nanshan and X-ray phases could 
be disregarded.

\subsubsection{Effect of the Proper Motion}

With accurate measurement by the Hubble Space Telescope, the proper motion
of the Crab pulsar has been obtained as $\mu_{\alpha}$= -11.8\,mas yr$^{-1}$ 
for R.A. and $\mu_{\delta}$= 4.4\,mas yr$^{-1}$ for decl. \citep{Kaplan2008}. 
However, the JBE uses a constant position for the Crab pulsar, and the influence 
of proper motion on the relative phases should be considered. If the pulsar 
position is changing, the timing residuals should show oscillations with 
gradually increasing amplitude \citep{Helfand1977}, which could be roughly 
described by $\Delta{\Phi}_{pm}=h\cdot\sin{\alpha}\cdot\Delta{\theta}/c_{0}\cdot{f}$,
where $h$ is the distance between the Earth and the Sun, $c_{0}$
is the speed of light, and $\alpha$ is the decl. of the Crab pulsar.
Taking into account the proper motion, the maximum value of the amplitude
is 0.0035\,periods for 10 years. However, the impact of proper motion is 
counteracted when the time is longer than one month, because the X-ray and 
Nanshan phases are the relative phases to JBE that were updated monthly. We 
check the power spectrum of these relative phases to see whether variation 
power exists on the time scale of about a month, which could be the impact 
of pulsar proper motion on the relative phases, and eventually no significant 
signals in the power spectrum of $\Phi_{P}$ and $\Phi_{H}$ have been found.
Thus, the proper motion is not the main reason for the long-term variation
of the X-ray and Nanshan phases. Similarly, the inaccuracy of solar system
ephemeris could not account for the long-term variation of the X-ray
phases.

\subsubsection{Check of the Spin Parameters}

Inaccurate spin parameters could lead to phase deviation. There are
some outliers in X-ray and Nanshan phases, especially the X-ray phases
around MJD 53350, which are obtained by using one ephemeris. These
results remind us to check the reliability of the JBE parameters, from 
the aspects of glitches, rotation frequencies, and radio reference TOAs.

Because the parameters of glitches have not been included in the JBE
directly \citep{Lyne(1993)}, we need to check whether the significant
residuals are caused by the glitches. However, we find that the obviously
abnormal phases are not coincident with the glitch epoches. The effects
of glitches after their occurrence month are greatly reduced. In MJD 
$53341$-$53372$, there is no glitch, so the outliers in this period did 
not resulted from glitches.

Furthermore, we check the parameters of JBE in MJD 53341--53372 in case
the JBE parameters are inaccurate. First, we compare the rotation 
frequencies inferred from JBE and those we searched from the three 
{\sl RXTE} observations in this period. The maximal difference between 
them is $(0.3\pm1.2)\times10^{-5}$ Hz for PCA and $(2.0\pm5.8)\times10^{-5}$ 
Hz for HEXTE, where the uncertainties are only from our frequency searching
process, which show that the frequencies we calculated from these three
observations are consistent with the JBE predictions. Second, we calculate 
the TOAs of these three X-ray observations using the frequencies we obtained 
above and the software TEMPO2 \citep{Edwards2006,Hobbs2006}, and then compare 
them with the TOAs inferred from the JBE time solution and the FTOOL command 
FASEBIN. As shown in Fig. \ref{fig1}, TOAs obtained with the above two methods 
are consistent. Therefore, both the JBE frequencies and the TOA calculation 
process are reliable, and the remaining possibility is that the reference TOA 
of JBE in MJD 53341--53372 is inaccurate, which causes the abnormal lags between
the X-ray and radio phases. We note that the inaccuracy of JBE TOAs has also 
been found by the Jodrell Bank Observatory, as on the web page of JBE it is 
pointed out, ``DO NOT trust the geocentric pulse arrival times 
yet!"\footnote{http://www.jb.man.ac.uk/pulsar/crab.html}.

\subsection{Is There a Significant Intrinsic Variation of the X-Ray
Emitting Region Relative to the Radio Ones?}

It is important for pulsar physics to find out whether the fluctuations 
and long-term evolutions of the X-ray phases are intrinsic, i.e., due 
to the relative geometric variations between the X-ray and radio emitting 
regions. We find that there are two observational facts contradicting 
with this hypothesis. As shown in Fig \ref{fig2}, the Nanshan radio 
phases are highly correlated with the X-ray phases derived from the PCA 
observations, which means that the two phases wander simultaneously. 
Furthermore, by using the Nanshan phases as a reference, the X-ray phases 
then have a smaller fluctuation amplitude and the long-term evolution 
disappears (Fig. \ref{fig3}c), which also implies that the X-ray and radio 
emitting regions do not have significant relative changes. Thus, the X-ray 
and radio phase fluctuations are both dominated by the pulsar spin.

\subsection{Constrains on timing noises}

The almost constant phase-lag between the X-ray and radio bands also 
supplies information about the origins of timing noises. Variations 
of the DM and thus the variations of ISM between the earth and pulsars 
have been detected, which can lead to the radio timing noises of those 
pulsars (You et al. 2007, and references there in). However, because 
the ISM has no effect on the X-ray TOAs, the constant value of the 
X-ray phase $\Phi_{P}^{'}$ means that most of the timing noises are 
not from the ISM, which is consistent with the results of PSR B1540$-$06 
obtained from the multi-frequency radio observations \citep{Hobbs(2010)}. 
Thus ISM variation cannot account for all the timing noises of the Crab pulsar.

\section{Summary}

Utilizing the 11-year X-ray observations from the RXTE, 6-year radio 
observations from Nanshan Telescope, and the ephemeris from the Jodrell 
Bank Observatory, we study the evolution of the X-ray and radio phases 
of the Crab Pulsar. The X-ray phases from PCA and HEXTE exhibit 
synchronous variations on all time scales, and X-ray and Nanshan phases 
also have a strong correlation with a Pearson's coefficient $r=0.78$. 
We find that the simultaneous secular changes and fluctuations of the 
X-ray phases $\Phi_{P}$, $\Phi_{H}$ and the Nanshan phases $\Phi_{N}$ 
are quite possibly caused by the unreliable reference TOAs in the JBE 
parameters. Using the Nanshan phases as timing reference, the corrected 
X-ray phases $\Phi_{P}^{'}$ show lower variation amplitude and remain 
almost constant over time with a change rate of $(-1.1\pm1.1)\times10^{-7}$\,period 
per day.

Based on the results above, we conclude that the distance of the X-ray
and radio emission regions on the Crab pulsar does not show detectable
secular changes, and the timing noises are not the result of any unique
physical processes in the radio or X-ray radiation regions. In addition,
the variation of the ISM is not the origin of Crab pulsar's timing noises,
which is consistent with the results obtained from the multi-frequency
radio observations of PSR B1540$-$06 \citep{Hobbs(2010)}.

\section*{Acknowledgments}
We appreciate Dr. Michael Smith, Lorenzo Natalucci, Craig Markwardt,
Yuanyue Pan, Liming Song, Jinlu Qu, Li Chen, and Jian Li for their
useful suggestions. This work is supported by the National Key Research
and Development Program of China (2016YFA0400802), National Science
Foundation of China (11233001, 11503027, and 11303069), the Strategic
Priority Research Program on Space Science, and the Chinese Academy of
Sciences, grant No. XDA04010300 and XDB23000000. This work is also 
partially supported by the Trainee Program of Qian Xuesen Laboratory 
of Space Technology. We thank the High Energy Astrophysics Science 
Archive Research Center (HEASARC) at NASA/Goddard Space Flight Center 
for maintaining its online archive service that provided the data used 
in this research.

%\clearpage

\clearpage

\clearpage

\clearpage

\clearpage

\clearpage

\clearpage

\begin{table}
\footnotesize
\caption {The change rates of X-Ray and radio phases}
\scriptsize{}
\label{table:1}
\medskip
\begin{center}
\begin{tabular}{c c c c c c c c c c c}

\hline \hline
& Instrument & MJD   & Energy Band & Change Rate        & Intercept $^{a}$  \\
&            &       &         & ($10^{-7}$ period/day) & ($10^{-3}$ period) \\
\hline
&   PCA        & 51955--55927  &2--60keV   & $5.0\pm0.9$   & $-8.8\pm0.1$ \\
&  HEXTE       & 51955--55927  &15--250keV & $4.5\pm0.9$   & $-8.7\pm0.1$   \\
\hline
&   PCA        & 53500--55693  &2--60keV   & $4.8\pm2.1$   & $-9.0\pm0.2$  \\
&   Nanshan    & 53500--55688  & 1540 MHz  & $6.3\pm1.0$   & $-1.6\pm1.0$\\
\hline
&  PCA$^{b}$ & 53500--55693  & ...        & $-1.1\pm1.1$  & $-7.4\pm0.1$  \\
\hline
&  HEXTE$^{c}$ & 51955--55927  &...       & $-0.2\pm0.3$  & $0.2\pm0.04$  \\
\hline
&PCA$^{d}$ & 50129--52941  &  2--16keV  &  $3.3\pm2.0$   & ...   \\
\hline
&PCA$^{e}$ & 51955--55142  &  2--60keV  &  $6.6\pm1.3$  & ...  \\
\hline \hline
\end{tabular}
\end{center}
{\bf Notes.}

$^a$ These intercepts correspond to the values at MJD 54000.

$^b$ The parameters for X-ray phases from PCA corrected by data of the Nanshan Telescope.

$^c$ The parameters for the phase lags between HEXTE and PCA.

$^d$ The result from \cite{Rots(2004)}.

$^e$ The result from \cite{Ge et al.(2012)}.
\end{table}

\begin{table}
\footnotesize
\caption {The correlation coefficients of the X-Ray and radio phases}
\scriptsize{}
\label{table:2}
\medskip
\begin{center}
\begin{tabular}{c c c c c c c}

\hline \hline
& Instrument & versus & Instrument & MJD   &   Slope          &$r$  \\
&            &    &            &       &                      & \\
\hline
&   PCA      & versus & HEXTE      & 51955-55927 &  $ 0.98\pm0.02$  & $0.96$ \\
&   PCA      & versus & Nanshan    & 53500-55693 &  $ 0.72\pm0.04$  & $0.78$ \\
\hline \hline
\end{tabular}
\end{center}
\end{table}

\end{document}